# It's Not the Size: Harness Design Determines Operational Stability in Small Language Models

Yong-eun Cho
KailosLab, Seoul, Republic of Korea
kevin@kailoslab.com

## Abstract

This paper experimentally analyzes how the level of harness engineering affects the operational performance of small language models (SLMs, 2–3B parameters). Three harness conditions—model-only (raw prompt), minimal-shell (wrapper tags), and a 4-stage pipeline (plan→execute→verify→recover)—are applied to three models (Gemma4 E2B, Qwen3.5:2B, LLaMA 3.2) across 24 tasks, comparing Task Success Rate (TSR) and Valid TSR (VTSR). The pipeline harness achieves TSR=0.952 and VTSR=1.000 on Gemma4 E2B (T1–T5, 21 tasks). A non-monotonic phenomenon—minimal-shell TSR < model-only TSR—is observed in two models. In LLaMA 3.2(3B) model-only, seven format violations yield TSR=0.429, revealing scaffold collapse: the model abandons JSON structure under complex format requirements without harness support. Ablation shows planning and recovery each contribute ~24.7% of total gain. VCR (Verification Catch Rate) = 0.625 across all pipeline runs.

## I. Introduction

The rapid development of LLMs has spurred growing interest in deploying small language models (2B–4B parameters) in edge or on-premises environments without cloud inference costs. However, most research focuses on static benchmark accuracy, leaving operational stability—output completeness and format compliance in multi-step tasks—largely unaddressed. Reflexion [1] and ReAct [2] showed that execution structure matters as much as raw capability, but systematic evidence for 2B–3B models is scarce.

This paper asks: can a well-designed execution harness enable small models to reliably achieve above-threshold performance within a constrained task scope? We answer empirically by applying three harness levels to three models across 24 tasks and measuring TSR/VTSR against pre-defined thresholds (TSR≥0.65, VTSR≥0.80).

Main contributions:

1) Systematic measurement of how harness complexity (none → minimal wrapper → 4-stage pipeline) affects SLM operational performance.

2) Discovery and characterization of a non-monotonic phenomenon: minimal harness yields lower TSR than no harness, in two models.

3) Evidence that the recovery mechanism is the primary contributor to pipeline gains, and classification of harness-fixable vs. unfixable failure modes.

4) Ablation showing the planning stage acts as a format anchor for quantitative constraints (e.g., character limits).

5) Introduction of scaffold collapse: under complex format requirements, LLaMA 3.2(3B) abandons JSON structure without harness support (TSR=0.429, 7 violations), showing harness format enforcement functions independently of content generation ability.

## II. Related Work

### 2.1 Execution Pipeline-based LLM Control

Reflexion [1] accumulates verbal feedback from failures for self-reflection in subsequent attempts. ReAct [2] interleaves reasoning and acting in an environment interaction loop. CRITIC [3] uses external tools to critically verify and revise outputs. These show that execution flow structure, not just raw model capability, significantly impacts performance. Our pipeline implements a lightweight self-correction loop without external tools, applicable to any Ollama-compatible model.

### 2.2 Small Language Model Research

Phi-3 [4], Gemma [5], Mistral [6], and Qwen [9] demonstrate that 2B–7B models can rival much larger models on specific tasks. However, these studies use static benchmarks (MMLU, HumanEval), leaving operational stability in multi-step scenarios unaddressed. Ollama [10] has made local SLM deployment practical, increasing the need for harness engineering research in this setting.

### 2.3 LLM Output Structuring Tools

LMQL [7], Guidance [8], and Outlines [11] enforce output structure via grammar constraints, improving format compliance but introducing tight runtime coupling. Our pipeline harness operates at the prompt level only—no grammar constraints, no runtime modifications—making it runtime-agnostic across all Ollama models.

## III. Experimental Design

### 3.1 Research Questions

RQ1: Can small models meet TSR≥0.65, VTSR≥0.80 within a constrained task scope?

RQ2: Does harness complexity monotonically benefit performance, or are non-linear effects present?

RQ3: Which pipeline component contributes most to performance gains?

RQ4: Are harness effects reproducible across different models?

RQ5: How do harness effects vary across task categories?

### 3.2 Experimental Setup

Models: Gemma4 E2B (gemma4:e2b, 2B), Qwen3.5:2B (qwen3.5:2b, 2B), LLaMA 3.2 (llama3.2:latest, 3B) via Ollama v0.21.2. Hardware: Windows 11, RTX 4060 8GB GPU. Inference: temperature=0.1, num_predict=2048; Gemma4/LLaMA num_ctx=8192, Qwen3.5 num_ctx=4096 (VRAM constraint). Timeout: 300s for 2B models; 600s for LLaMA 3.2(3B) on dedicated GPU. Each task is run once; retries occur only within the pipeline recovery stage. Duplicate runs: only the most recent result is included in analysis.

### 3.3 Harness Conditions

(A) model-only: Raw prompt with task description only. No structural wrapper.

(B) minimal-shell: Task wrapped in [TASK START] / [OUTPUT] tags. No execution flow.

(C) 4-stage pipeline: Plan → Execute → Verify → Recover. Recovery re-runs (up to 2 retries) with error feedback on verify failure or timeout.

### 3.4 Ablation Conditions

**Table I. Ablation condition definitions**

| Condition | Removed Stage | Remaining Stages |
|---|---|---|
| pipeline-no-plan | Planning | Execute → Verify → Recover |
| pipeline-no-verify | Verify + Recover | Plan → Execute |
| pipeline-no-recover | Recover | Plan → Execute → Verify (verdict only) |

### 3.5 Task Design

24 tasks across 6 categories are designed to span common SLM operational scenarios. T6 (Web Search) uses DuckDuckGo API in the pipeline condition only; model-only and minimal-shell lack tool access.

**Table II. Task category design**

| Category | Description | Tasks | Web Search |
|---|---|---|---|
| T1: Struct. Knowledge | Structured information extraction & organization | 4 | × |

| T2: Search-and-Ground | Document-based search with evidence grounding | 4 | × |
|---|---|---|---|
| T3: Comparison | Cross-item comparative analysis | 4 | × |
| T4: Workflow Completion | Multi-step procedure execution | 4 | × |
| T5: Constraint-Sensitive | Quantitative and format constraint compliance | 5 | × |
| T6: Web Search Task | Web search-based response generation | 3 | ✓ (pipeline only) |

### 3.6 Evaluation Metrics

All outputs are manually scored on a 0/1/2 rubric (rubric.md). TSR (Task Success Rate) = #score-2 tasks / total tasks. VTSR (Valid TSR) = #score≥1 tasks / total tasks. VCR (Verification Catch Rate) = #(retry_count>0 among score<2) / #score<2. Thresholds TSR≥0.65 and VTSR≥0.80 are jointly required for 'stably operable'. TSR 0.65 requires full success in at least two-thirds of tasks; VTSR 0.80 ensures fewer than 20% complete failures (score=0).

### 3.7 Reproducibility

Tasks are fixed in task_inputs/*.json (ID, instruction, input data, expected format). Conditions are run sequentially (single batch script) to avoid Ollama queue cascades. Each run is saved to results/{condition}/{run_id}.json with full metadata (model, condition, timestamp, elapsed time, retry count, pipeline trace). All prompts, task definitions, result files, and analysis code are version-controlled.

## IV. Results — Gemma4 E2B Baseline Experiment

### 4.1 Overall Performance (T1–T5, 21 tasks)

T6 is excluded from baseline comparison: model-only and minimal-shell have no tool access, making T6 score=0 structural rather than informative. All three conditions meet both thresholds (positive answer to RQ1). 2B parameter models can achieve above-threshold performance within a constrained task scope.

**Table III. Harness condition performance (Gemma4 E2B, T1–T5, 21 tasks)**

| Condition | TSR | VTSR | Threshold Met |
|---|---|---|---|
| model-only | 0.762 (16/21) | 0.857 (18/21) | ✓ |
| minimal-shell | 0.714 (15/21) | 0.810 (17/21) | ✓ |
| **Pipeline** | **0.952 (20/21)** | **1.000 (21/21)** | ✓ |

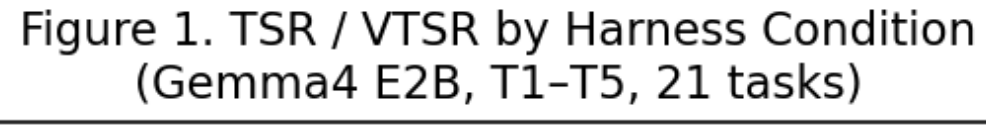

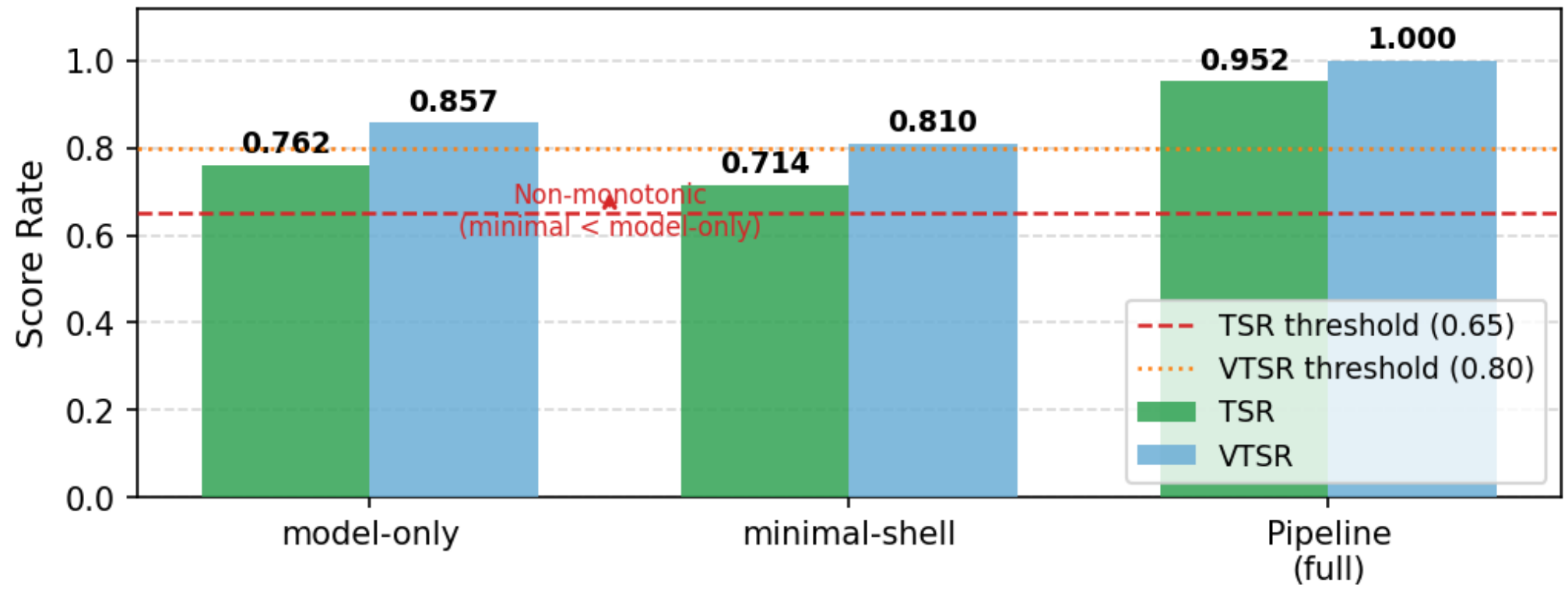

*Figure 1. TSR/VTSR comparison by harness condition (Gemma4 E2B, T1–T5). Pipeline achieves TSR=0.952, VTSR=1.000. Note the non-monotonic reversal: minimal-shell TSR (0.714) < model-only (0.762).*

### 4.2 Condition-level Analysis

model-only vs. minimal-shell reversal: wrapper tags are estimated to increase cognitive load in complex workflow tasks (T4), inducing additional timeouts. model-only had 3 timeouts (T1-03, T4-01, T4-03); minimal-shell added T4-04 (4 total). Note: cognitive load was not directly measured; controlled experiments are needed.

Pipeline superiority: TSR=0.952, VTSR=1.000. The recovery mechanism is the primary contributor—retrying T1-03, T4-01, T4-03 (timed out in other conditions) and converting all to success. The sole unsolved task is T5-04 (score=1, 542-char output vs. 500-char limit). Pipeline incurs ~2–3× inference overhead vs. model-only.

### 4.3 Task Category Analysis

**Table IV. Category-level TSR comparison (Gemma4 E2B)**

| Category | model-only | minimal-shell | Pipeline | Max Gain |
|---|---|---|---|---|
| T1: Struct. Know. | 0.75 (3/4) | 0.75 (3/4) | 1.00 (4/4) | +0.25 |
| T2: Search-Ground | 1.00 (4/4) | 1.00 (4/4) | 1.00 (4/4) | 0 |
| T3: Comparison | 1.00 (4/4) | 1.00 (4/4) | 1.00 (4/4) | 0 |
| T4: Workflow | 0.50 (2/4) | 0.25 (1/4) | **1.00 (4/4)** | +0.75 |
| T5: Constraint | 0.60 (3/5) | 0.60 (3/5) | 0.80 (4/5) | +0.20 |

** T4 shows largest pipeline gain (+0.75 from minimal-shell baseline).*

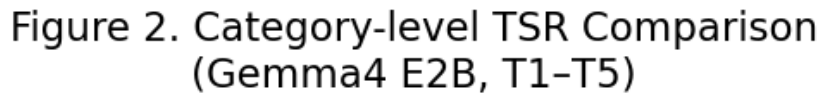

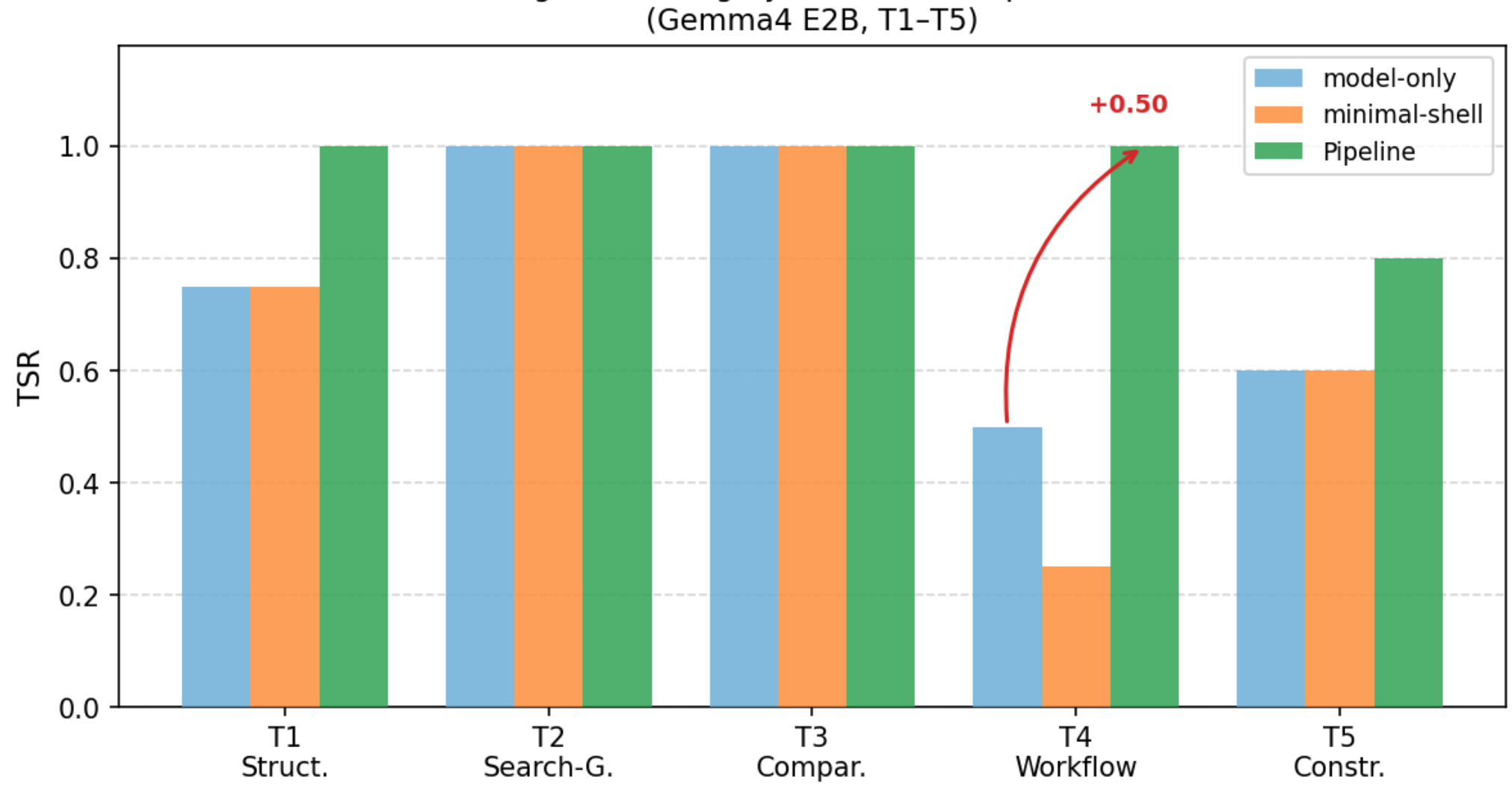

*Figure 2. Category-level TSR comparison (Gemma4 E2B, T1–T5). T2 and T3 achieve TSR=1.00 under all conditions. T4 shows the largest pipeline gain (+0.50 from model-only, +0.75 from minimal-shell).*

T4 (Workflow Completion): Pipeline TSR=1.00 vs. model-only 0.50 / minimal-shell 0.25. Pipeline planning pre-structures execution flow; recovery handles timeout retries. T2 and T3: all conditions TSR=1.00—structurally simple tasks with provided documents. T5: pipeline 0.60→0.80 by fixing T5-02 (prohibited word 'burakdeurimnida'). T5-04 (500-char limit) fails across all conditions—LLM character-count estimation limitations prevent correction even via verify-recover.

### 4.4 Failure Mode Analysis

T5-04 (500-character report) scores 1 in all conditions, exemplifying the harness-unfixable failure category. Failure modes are classified by post-hoc (reactive) harness remediation feasibility.

**Table V. Failure mode classification and harness remediation**

| Failure Mode | Description | Harness Fixable |
|---|---|---|
| incomplete_completion | Timeout or empty output | ✓ (recovery retry) |
| format_violation | Required output format not followed | △ (verify-recover loop) |
| grounding_failure | Uses information beyond provided docs | △ |
| missing_step | Multi-step procedure partially omitted | △ |
| constraint_violation | Quantitative constraint violated | × (model limit) |

### 4.5 Verification Catch Rate (VCR)

Across all pipeline runs (3 models, T1–T6): denominator = 16 score<2 tasks (Gemma 4 + Qwen 5 + LLaMA 7); numerator = 10 tasks with retry_count>0. VCR = 10/16 = 0.625. Model breakdown: Gemma 3/4=0.750, LLaMA 7/7=1.000, Qwen 0/5=0.000. LLaMA verification catches all defects but cannot fix them; Qwen score<2 cases are all false passes. Gemma T1–T5 VCR = 0/1 = 0.000: T5-04 is a false pass (verifier miscounts characters as ≤500).

### 4.6 T6 Web Search Task Analysis

**Table VI. T6 Web Search Task results**

| Condition | T6 TSR | Note |
|---|---|---|
| model-only | 0.000 (0/3) | No tool access — structural failure |
| minimal-shell | 0.000 (0/3) | No tool access — structural failure |
| Pipeline | 0.000 (0/3) | DuckDuckGo rate-limit (HTTP 202) all 3 runs |

T6 design limitation: model-only and minimal-shell structurally score=0 because they lack tool access. This tests 'tool-equipped vs. tool-less' rather than 'orchestration quality'. A valid design would give all conditions identical tool access and vary orchestration level only. T6 results confirm the functional existence of external tool integration, not orchestration superiority. T6-inclusive pipeline TSR: 0.833 (20/24).

## V. Results — Cross-Model Comparison

### 5.1 Cross-Model Performance (T1–T5 and T1–T6)

The same three harness conditions are applied to Qwen3.5:2B and LLaMA 3.2 to test harness effect reproducibility (RQ4). T1–T5 and T1–T6 metrics are reported separately because T6 structurally disadvantages model-only/minimal-shell.

**Table VII. Cross-model performance comparison (T1–T6, 24 tasks)**

| Model | Condition | TSR (T1-T5) | VTSR (T1-T5) | TSR (T1-T6) | VTSR (T1-T6) | Met† |
|---|---|---|---|---|---|---|
| Gemma4 E2B | model-only | 0.762 (16/21) | 0.857 (18/21) | 0.667 (16/24) | 0.750 (18/24) | T1-T5 ✓ |
| Gemma4 E2B | minimal-shell | 0.714 (15/21) | 0.810 (17/21) | 0.625 (15/24) | 0.708 (17/24) | T1-T5 ✓ |
| Gemma4 E2B | Pipeline | 0.952 (20/21) | 1.000 (21/21) | 0.833 (20/24) | 0.875 (21/24) | ✓ |
| Qwen3.5:2B | model-only | 0.952 (20/21) | 1.000 (21/21) | 0.833 (20/24) | 0.875 (21/24) | ✓ |
| Qwen3.5:2B | minimal-shell | 0.857 (18/21) | 1.000 (21/21) | 0.750 (18/24) | 0.875 (21/24) | ✓ |
| Qwen3.5:2B | Pipeline | 0.857 (18/21) | 1.000 (21/21) | 0.792 (19/24) | 0.958 (23/24) | ✓ |
| LLaMA 3.2(3B) | model-only‡ | 0.429 (9/21) | 0.952 (20/21) | 0.375 (9/24) | 0.833 (20/24) | × |
| LLaMA 3.2(3B) | minimal-shell§ | 0.810 (17/21) | 1.000 (21/21) | 0.708 (17/24) | 0.875 (21/24) | ✓ |
| LLaMA 3.2(3B) | Pipeline§ | 0.762 (16/21) | 0.857 (18/21) | 0.708 (17/24) | 0.875 (21/24) | ✓ |

*† Threshold: TSR≥0.65 and VTSR≥0.80 (T1–T5 basis). ‡ LLaMA model-only: timeout=600s, dedicated GPU; TSR failure = 7 format violations. § LLaMA minimal-shell/pipeline: separate session (different GPU load) — direct comparison indicative only.*

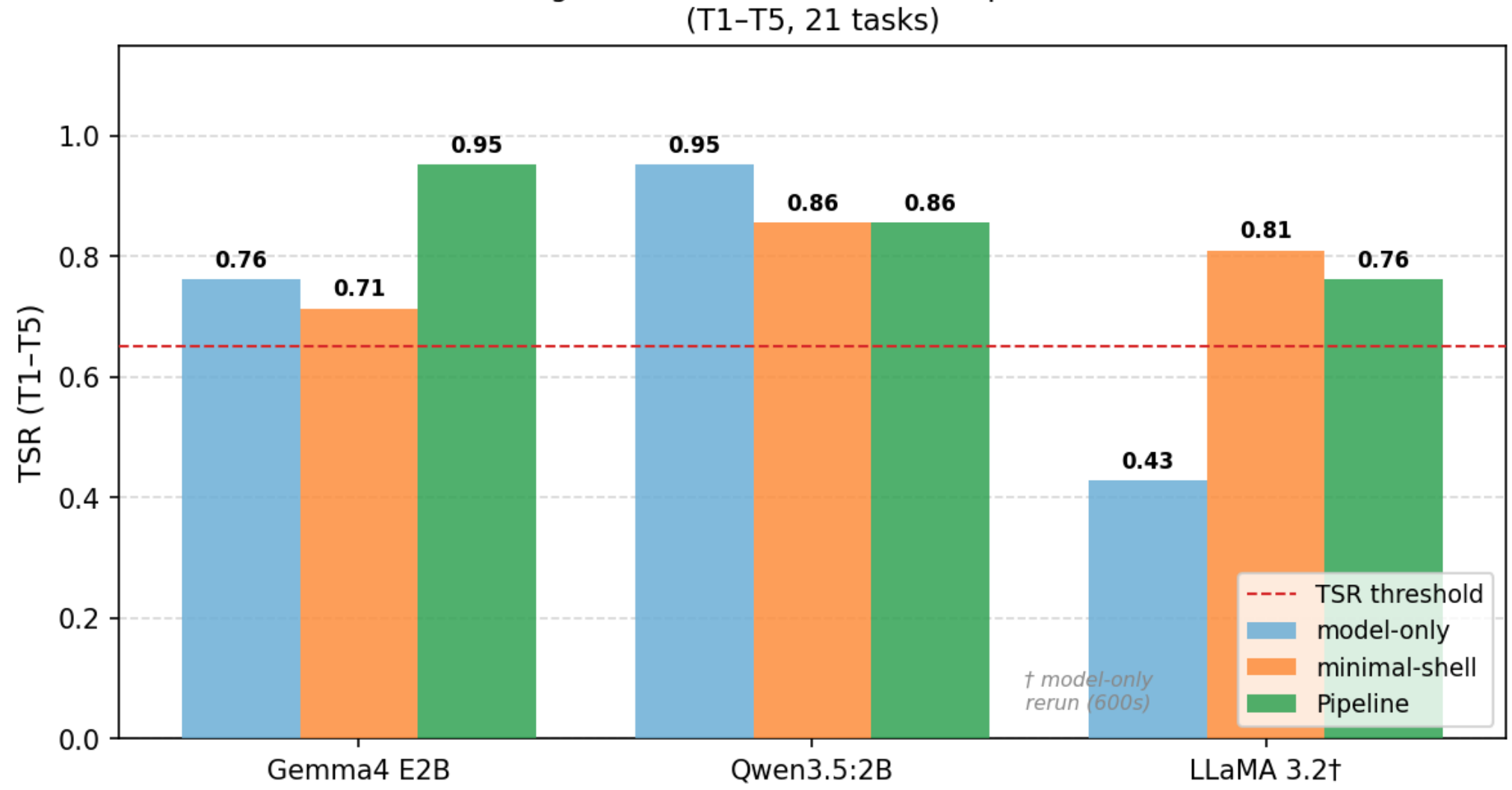

*Figure 4. Cross-model TSR comparison (T1–T5). Gemma4 E2B shows largest pipeline gain (+0.190). Qwen3.5:2B pipeline TSR (0.857) falls below model-only (0.952) — verify-recover induces repeated score=1 in T2. LLaMA 3.2(3B) model-only (TSR=0.429) fails threshold without harness.*

## 5.2 Task × Model Score Matrix (Pipeline Condition)

**Table VIII. Task × model score comparison (pipeline condition)§**

| **Task** | **Category** | **Gemma4 E2B** | **Qwen3.5:2B** | **LLaMA 3.2** |
|---|---|---|---|---|
| T1-01 | Struct. Know. | 2 | 2 | 2 |
| T1-02 | Struct. Know. | 2 | 2 | 0 |
| T1-03 | Struct. Know. | 2 | 2 | 2 |
| T1-04 | Struct. Know. | 2 | 2 | 1 |
| T2-01 | Search-Ground | 2 | 2 | 2 |
| T2-02 | Search-Ground | 2 | 1 | 0 |
| T2-03 | Search-Ground | 2 | 1 | 1 |
| T2-04 | Search-Ground | 2 | 1 | 0 |
| T3-01 | Comparison | 2 | 2 | 2 |
| T3-02 | Comparison | 2 | 2 | 2 |
| T3-03 | Comparison | 2 | 2 | 2 |
| T3-04 | Comparison | 2 | 2 | 2 |
| T4-01 | Workflow | 2 | 2 | 2 |
| T4-02 | Workflow | 2 | 2 | 2 |
| T4-03 | Workflow | 2 | 2 | 2 |
| T4-04 | Workflow | 2 | 2 | 2 |
| T5-01 | Constraint | 2 | 2 | 2 |
| T5-02 | Constraint | 2 | 2 | 2 |
| T5-03 | Constraint | 2 | 2 | 2 |
| T5-04 | Constraint | 1 | 2 | 2 |
| T5-05 | Constraint | 2 | 2 | 2 |
| T6-01 | Web Search | 0 | 0 | 1 |
| T6-02 | Web Search | 0 | 2 | 2 |
| T6-03 | Web Search | 0 | 1 | 1 |
| **TSR (T1-T5)** | | **0.952** | **0.857** | **0.762** |
| **VTSR (T1-T5)** | | **1.000** | **1.000** | **0.857** |

*§ LLaMA minimal-shell and pipeline scores from separate session—see §7.5 limitation 6.*

## 5.3 Model × Condition Pattern Analysis

Gemma4 E2B: model-only TSR=0.762 → pipeline 0.952 (+0.190). Gain attributable to timeout recovery (T1-03, T4-01, T4-03). Qwen3.5:2B: model-only TSR=0.952 (already high—fast inference ~15s/task avoids timeouts). Pipeline TSR=0.857 (−0.095 reversal) because verify-recover loop in T2 repeatedly generates score=1 outputs (structurally correct but incomplete citation). Non-monotonic phenomenon confirmed in both Gemma and Qwen (T1–T5 basis): Gemma minimal-shell 0.714 < model-only 0.762; Qwen 0.857 < 0.952.

LLaMA 3.2(3B) model-only completed all tasks in 4–21s (fast inference). TSR=0.429, VTSR=0.952: content generation ability present, format stability absent. Category TSR details are shown in Table IX below.

**Table IX. LLaMA 3.2(3B) model-only category performance (T1–T5)**

| Category | TSR | VTSR | Note |
|---|---|---|---|
| T1: Struct. Know. | 0.250 (1/4) | 1.000 | JSON array errors (3 tasks) |
| T2: Search-Ground | 0.500 (2/4) | 1.000 | Hallucination 1, missing_step 1 |
| T3: Comparison | 1.000 (4/4) | 1.000 | Full success |
| T4: Workflow | 0.250 (1/4) | 0.750 | Markdown output 1 (score=0) |
| T5: Constraint | 0.200 (1/5) | 1.000 | Format violations 4 |
| **Total** | **0.429 (9/21)** | **0.952 (20/21)** | |

# VI. Ablation Study (Gemma4 E2B, T1–T6)

## 6.1 Pipeline Component Contribution

**Table X. Ablation condition performance (Gemma4 E2B, T1–T6, 24 tasks)**

| Condition | Removed Stage | TSR (T1-T6) | VTSR (T1-T6) | Met |
|---|---|---|---|---|
| **Pipeline (full)** | — | **0.833** (20/24) | **0.875** (21/24) | ✓ |
| pipeline-no-plan | Planning | 0.792 (19/24) | 0.875 (21/24) | ✓ |
| pipeline-no-verify† | Verify + Recover | 0.909 (20/22) | 1.000 (22/22) | ✓ |
| pipeline-no-recover | Recover | 0.792 (19/24) | 0.875 (21/24) | ✓ |
| model-only (baseline) | All stages | 0.667 (16/24) | 0.750 (18/24) | × |

*† 2 tasks unscored (run error); scored on 22 tasks. TSR reversal explained in §6.3.*

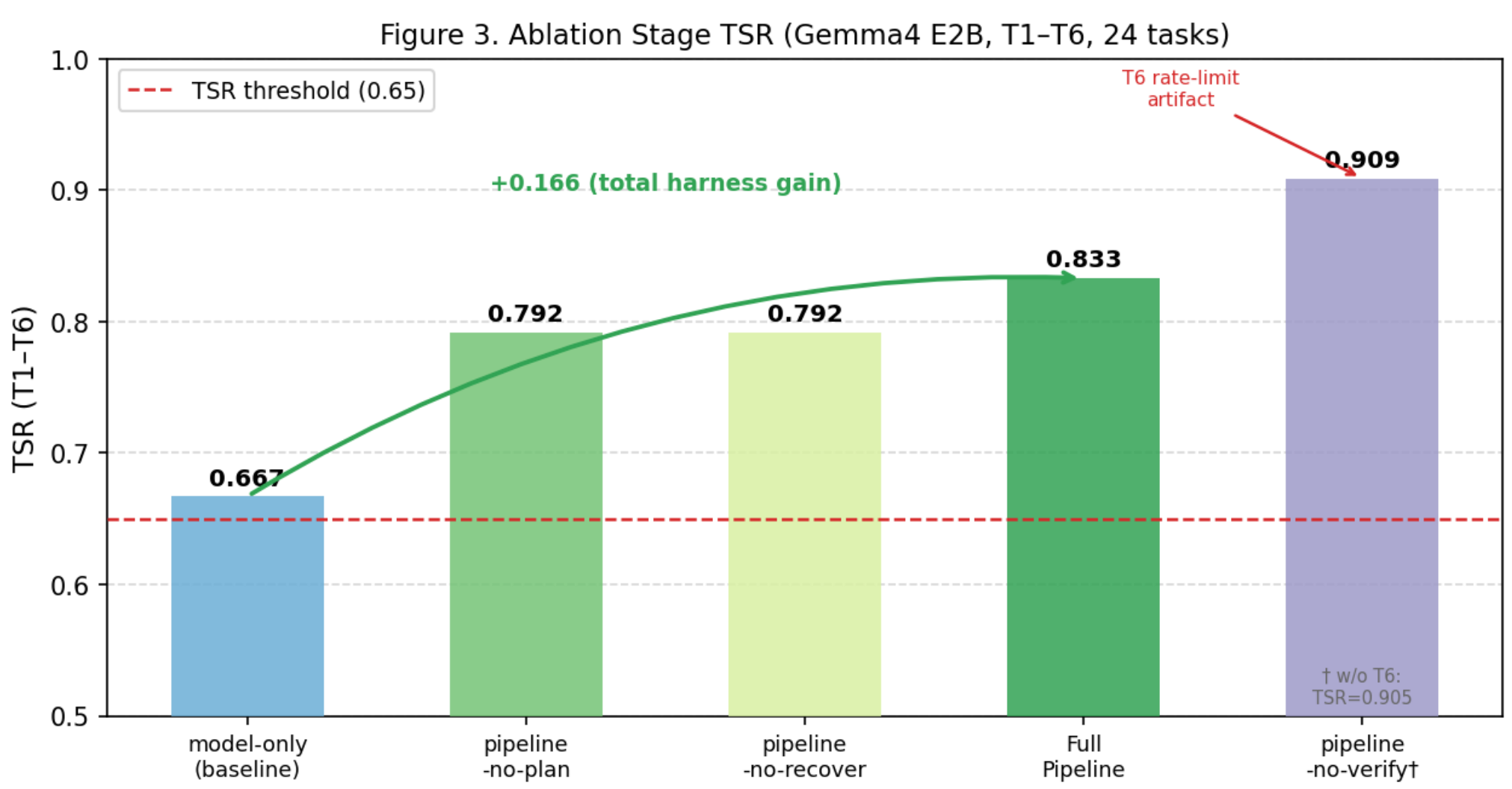

*Figure 3. Ablation stage TSR (Gemma4 E2B, T1–T6). model-only (0.667) → full pipeline (0.833): total gain +0.166. pipeline-no-verify TSR=0.909 reversal is a T6 rate-limit artifact (§6.3).*

**Table XI. Pipeline component contribution analysis**

| Removed Stage | Resulting TSR | TSR Drop vs. Full | % of Total Gain |
|---|---|---|---|
| Planning | 0.792 | −0.041 | 24.7% |
| Verify + Recover | 0.909 | −0.076 (reversed†) | N/A |
| Recover | 0.792 | −0.041 | 24.7% |

*† pipeline-no-verify TSR > full pipeline: T6 rate-limit artifact, not genuine reversal (§6.3).*

VCR analysis: full pipeline VCR=0.625 (10/16). pipeline-no-verify: VCR=0.0 (no verify/recover stages). pipeline-no-recover: VCR=0.0 (verify detects but no retry → retry_count=0 → numerator=0).

### 6.2 Planning Stage as Format Anchor

**Table XII. Category TSR: full pipeline vs. pipeline-no-plan (Gemma4 E2B)**

| Category | Full Pipeline | No-Plan Pipeline | Difference |
|---|---|---|---|
| T1 | 1.000 (4/4) | 0.750 (3/4) | **−0.250** |
| T2 | 1.000 (4/4) | 1.000 (4/4) | 0 |
| T3 | 1.000 (4/4) | 1.000 (4/4) | 0 |
| T4 | 1.000 (4/4) | 1.000 (4/4) | 0 |
| T5 | 0.800 (4/5) | 0.800 (4/5) | 0 |
| T6 | 0.000 (0/3) | 0.000 (0/3) | 0 |
| **Total** | **0.833** | **0.792** | **−0.041** |

The T1–T5 difference between full pipeline and no-plan is entirely due to T1-04 (200-character summary). Without planning: 248-char output. With planning: the plan step explicitly decomposes '200-char limit' as an execution anchor → 193-char output. This demonstrates planning as a proactive quantitative constraint anchor, not just a task decomposition step.

### 6.3 Verify+Recover Reversal: T6 Rate-Limit Artifact

pipeline-no-verify TSR=0.909 > full pipeline TSR=0.833 appears counterintuitive. Root cause: T6-01 succeeded in the no-verify run (different execution timing avoided the DuckDuckGo rate limit; full pipeline T6 all 3 runs hit rate-limit). This 1-task difference accounts for 1/24 ≈ 0.042 TSR gap. Excluding T6: full pipeline T1-T5 TSR=0.952 > no-verify T1-T5 TSR=0.905, restoring expected ordering. The reversal is a T6 rate-limit noise artifact.

## VII. Discussion

### 7.1 Harness as Scaffold, Not Just Retry

LLaMA 3.2(3B) model-only results reveal two independent harness roles: (1) reactive recovery—retrying timed-out tasks; (2) proactive scaffolding—enforcing output format structure before execution. VTSR=0.952 confirms content generation ability; TSR=0.429 reflects format collapse. These two roles address distinct failure types and are not interchangeable.

### 7.2 Non-Monotonic Harness Effect

Gemma and Qwen both show minimal-shell TSR < model-only TSR (T1–T5 basis). The pattern is consistent with wrapper tags increasing cognitive load, causing additional timeouts in complex tasks—but this is hypothesis, not proven causation. Controlled experiments isolating prompt length vs. structural complexity are needed. 'Less can be more': half-measures in harness design may hurt rather than help.

### 7.3 Failure Mode Classification (Extended)

**Table XIII. Extended failure type classification and harness mechanisms**

| Failure Type | Harness Fixable | Mechanism |
|---|---|---|
| Timeout (incomplete_completion) | ✓ (reactive) | Recovery retry (if timeout budget sufficient) |
| Rule-detectable constraint (e.g., prohibited words) | ✓ (reactive) | Verify → recover loop |
| Scaffold collapse (format abandonment) | ✓ (proactive) | Planning + prompt structuring (pre-emptive) |
| Quantitative constraint (char count) | △ | LLM counting limit → rule-based verification needed |
| Hallucination (knowledge boundary) | × | Model capability limit |

### 7.4 T6 Design Limitation and Redesign Proposal

Current T6 design compares 'tool-equipped vs. tool-less', confirming only that having a tool beats not having one—a trivial result for harness research. A correct design for measuring orchestration quality:

**Table XIV. Proposed T6 equal-tool redesign**

| Design Condition | Tool Access | Orchestration Level |
|---|---|---|
| tool-only (baseline) | Identical search API | Raw model call, no structure |
| tool-with-plan | Identical search API | Planning stage formulates search strategy |
| tool-with-pipeline | Identical search API | 4-stage: search → execute → verify → recover |

This design measures 'how reliably the same tool is used' as a function of orchestration level. T6 results in this paper should be treated as confirming the functional existence of external tool integration, not orchestration superiority.

### 7.5 Limitations

1) Small task set (24 tasks) limits generalization beyond tested domains and difficulty levels.

2) Single run per task—no variance estimation or confidence intervals.

3) Non-uniform timeout (300s vs. 600s) affects timeout-based comparisons across models.

4) Single scorer—no inter-rater reliability measurement.

5) Pipeline overhead: ~2–3× inference time vs. model-only (latency-sensitive applications).

6) LLaMA cross-condition inconsistency: model-only run on dedicated GPU; minimal-shell/pipeline from separate session (different load) — comparison is indicative only.

7) External service dependency (T6): DuckDuckGo rate-limits caused all pipeline T6 runs to fail.

## VIII. Conclusion

We experimentally demonstrated that within our experimental scope (3 models, 24 tasks), SLM operational stability depends more on harness design than on parameter count. All three conditions met operational thresholds for Gemma4 E2B (T1–T5); the pipeline achieves TSR=0.952, VTSR=1.000. Key findings:

RQ1 (feasibility): SLMs can meet pre-defined operational thresholds within constrained scope.

RQ2 (non-monotonic): minimal harness can reduce TSR below no-harness (Gemma and Qwen).

RQ3 (components): planning and recovery each contribute ~24.7% of total gain; VCR=0.625.

RQ4 (reproducibility): harness effect confirmed in Qwen; but gain pattern differs by model. LLaMA 3.2(3B) model-only (TSR=0.429) confirms harness necessity for format-unstable models.

RQ5 (categories): T4 (Workflow) shows largest gain; T2–T3 need no harness; T5 partially fixable; T6 design flawed.

Core message: harness performs two independent functions—reactive recovery and proactive scaffolding—that address distinct failure types. Future work: (1) rule-based verification (regex, Python counters) to improve VCR; (2) diversified recovery strategies; (3) unified re-measurement with identical environments for all model-condition pairs; (4) equal-tool T6 redesign; (5) lightweight pipeline for latency-sensitive applications.